\documentclass[12pt]{iopart}

\newcommand{\om}{\Omega_{\rm M}}
\newcommand{\ola}{\Omega_{\rm\Lambda}}
\usepackage{iopams}
\usepackage{epsfig} 

\begin{document}

\title{Limiting the dimming of distant type Ia supernovae}
\author{Linda \"{O}stman\dag and Edvard M\"{o}rtsell\ddag}
\address{\dag\ Department of Physics, Stockholm University,
         SE 106 91 Stockholm, Sweden}
\address{\ddag\ Department of Astronomy, Stockholm University,
         SE 106 91 Stockholm, Sweden}

\eads{\mailto{\dag\ linda@physto.se}}
\eads{\mailto{\ddag\ edvard@astro.su.se}}

\begin{abstract}

Distant supernovae have been observed to be fainter than what is
expected in a matter dominated universe. The most likely explanation
is that the universe is dominated by an energy component with negative
pressure -- dark energy. However, there are several astrophysical
processes that could, in principle, affect the measurements and in
order to be able to take advantage of the growing supernova
statistics, the control of systematic effects is crucial. We discuss
two of these; extinction due to intergalactic grey dust and dimming
due to photon-axion oscillations and show how their effect on
supernova observations can be constrained using observed quasar
colours and spectra. For a wide range of intergalactic dust models, we
are able to rule out any dimming larger than 0.2 magnitudes for a type
Ia supernova at $z=1$. The corresponding limit for intergalactic Milky
Way type dust is 0.03 mag. For the more speculative model of photons
mixing with axions, we find that the effect is independent of photon
energy for certain combinations of parameter values and a dimming as
large as 0.6 magnitudes cannot be ruled out. These effects can have
profound implications for the possibility of constraining dark energy
properties using supernova observations.
 
\end{abstract}



\section{Introduction}

Observations of distant type Ia supernovae (SNIa) show that they are
fainter than what the theory predicts for a universe described by a
standard flat cold dark matter model. The most popular explanation is
that the universe not only consists of luminous and dark matter, but
also of large amounts of dark energy, a generic name for a
homogeneously distributed energy component with negative pressure,
$p<-\rho /3$\footnote{We do not consider any fluctuations of the dark
energy component in this paper.}. This is also verified by other
cosmological probes such as the power spectra obtained from the
temperature fluctuations in the cosmic microwave background
\cite{Bennett:2003bz,Spergel:2003cb} and (on smaller scales) from
large scale galaxy surveys
\cite{2dFGRS,Tegmark:2003ud,Tegmark:2003uf}.

Several mechanisms have been suggested in order to explain the
faintness of distant SNIa without having to introduce a dark energy
component. Since the existence of a large dark energy component has
been independently detected by a number of cosmological probes, it is
unlikely that all the dimming is caused by dust or photon-axion
oscillations. However, it is crucial to control all possible
systematic effects in current and future SN surveys focussing on
constraining the time evolution of the dark energy component.

{\em Supernova evolution:} Some of the apparent dimming of distant
SNIa could be due to the fact that their luminosities are redshift
dependent and that the SNe thus cannot be used as standard
candles. This could, e.g., be caused by evolution of the SN progenitor
\cite{Dominguez:1998jt,Riess:1998cb}.
Evolutionary effects can be searched for by comparing the light curve
shape and spectra of SNe at different redshifts and
environments. Most studies show that the luminosity properties are
rather similar \cite{Perlmutter:1998np,Sullivan:2002ub}, but others
have found variations indicating evolutionary effects
\cite{Dominguez:1998jt,Hoeflich:1998uk,Riess:1999ti}. 

{\em Gravitational lensing:} Since matter is inhomogeneously
distributed in the universe, most line of sights to distant objects
will probe a matter density significantly lower than the average
density and the corresponding images will be demagnified relative to a
homogeneous universe. It is generally accepted that gravitational
lensing cannot be responsible for all the observed dimming of distant
SNIa, however the effect may still lead to a non-negligible bias that
need to be corrected for, e.g., by using information from the
magnitude distribution \cite{Amanullah:2002xh}. It has also been
suggested that the large scale expansion of the universe can be
affected by the inhomogeneities in the matter distribution. However,
perturbative calculations show that this effect is probably negligible
\cite{Kolb:2004,Rasanen:2004}.

{\em Photon-axion oscillations:} It has been proposed that the photon
flux from SNIa may be diminished because of photons oscillating into
axions in much the same way as the electron neutrino ($\nu_e$) flux,
produced in the sun, is diminished because of oscillations into muon
neutrinos ($\nu_{\mu}$) and tau neutrinos ($\nu_{\tau}$). Due to
differences in methods and assumptions regarding the magnetic field
strengths and plasma densities, different groups have reached
different conclusions about the possible magnitude and the energy
dependence of the effect
\cite{Csaki:2001yk,Deffayet:2001pc,Csaki:2001jk,Mortsell:2002dd,
Christensson:2002ig,Csaki:2004}.

{\em Intergalactic grey dust:} It has been suggested
that intergalactic dust could explain the observed faintness of high
redshift SNIa \cite{Aguirre:1998ge, Aguirre:1999dc}. If there is dust
along the line of sight, it would extinguish some of the SN light
through absorption and scattering. The intergalactic dust has to be
``grey'' in order not to cause a significant reddening that would have
been discovered if it existed.

In this paper, we are going to discuss intergalactic dust extinction
and photon-axion oscillations and show how the effects can be
constrained using quasars in the first data release (DR1) from the
Sloan Digital Sky Survey (SDSS). Similar studies have been performed
using quasars from the SDSS early data release (EDR) for intergalactic
dust \cite{Mortsell:2003vk} and for photon-axion oscillations
\cite{Mortsell:2003ya}. For intergalactic dust, a maximum allowed
amount of dimming of 0.2 magnitudes for SNIa at $z=1$ was found. In
this paper we redo the analysis using the larger and higher quality
data set of DR1. We also extend the range of dust models studied. In
the analysis of photon-axion oscillations, there was an erroneous
factor involving the photon energy in the expression for the density
matrix evolution. We redo this analysis using the larger data set of
DR1 with the correct expression for the density matrix evolution.

In Section \ref{sec:oscillations} we constrain the effects from
photon-axion oscillations and in Section \ref{sec:dust} we discuss limits
on the intergalactic dust attenuation. In Section \ref{sec:summary} we
summarize our results.

\section{Photon-axion oscillations}
\label{sec:oscillations}

The axion is a hypothetical neutral boson with spin zero originally
introduced as part of a possible solution to the strong charge-parity
(CP) violation. It has also been proposed as a candidate for the
missing dark matter in the universe. An axion can decay into two
photons, $a \to 2 \gamma$ but also oscillate into a photon. Since they
have different spin\footnote{The axion has spin zero while the photon
has spin one.}, photons and axions can only mix in the presence of a
mixing agent that preserves the quantum numbers such as angular
momentum, e.g. an external transverse magnetic or electric field
\cite{Raffelt:1996wa}.

The photon axion interaction is described by the Lagrangian
\begin{equation}
	{\cal L}_{\rm int}=\frac{a}{M_{\rm a}}\vec E\cdot\vec B ,
\end{equation}
where $a$ is the axion field, $M_{\rm a}$ is a mass scale and $\vec
E$
 and $\vec B$ are the electric and magnetic field,
respectively. The
 mass scale determines the strength of the
coupling. Recent results
 from the CERN Axion Solar Telescope (CAST)
implies $M_{\rm
 a}>8.62\cdot 10^{9}$ GeV at 95\% confidence level for
an axion mass
 $m_{\rm a}\lesssim 0.02$ eV \cite{Andriamonje:2004}.
 
\subsection{Theory of the simulations}

To determine the amount of dimming different scenarios of photon-axion
oscillations lead to, we calculate the conversion probability for
oscillations using the density matrix formalism. The evolution of a
density matrix $\rho$, is described by \cite{Sakurai:1995aa}
\begin{equation}
\label{eq:ih}
  {\rm i}\hbar \frac{\partial \rho}{\partial t} = -[\rho,{\cal H}],
\end{equation}
where ${\cal H}$ is the Hamiltonian, which in our case is given by
${\cal H} = \frac{\hbar}{2} M$. The mixing matrix $M$ is given by
\cite{Deffayet:2001pc,Goobar:2002vm}
\begin{equation}
  \label{eq:M} 
  M=\left(\begin{array}{ccc}
       \Delta_{\perp} & 0 & \Delta_{\rm M}\cos\alpha\\
       0 & \Delta_{\parallel} & \Delta_{\rm M}\sin\alpha\\
       \Delta_{\rm M}\cos\alpha & \Delta_{\rm M}\sin\alpha & \Delta_{\rm m}
       \end{array}\right),
\end{equation}
where $\alpha$ is the angle in the plane perpendicular to the
propagation between a fixed polarisation vector and the projected
magnetic field. The other quantities in the mixing matrix are
\begin{eqnarray}
  \label{eq:terms} 
\Delta_{\perp} & = & -3.6 \cdot 10^{-25}\left(\frac{\omega}{1\,{\rm eV}}\right)^{-1}
\left(\frac{n_{\rm e}}{10^{-8}\,{\rm cm}^{-3}}\right){\rm cm}^{-1},\nonumber \\ 
\Delta_{\parallel} & = & \Delta_{\perp},\nonumber \\
\Delta_{\rm M} & = & 2 \cdot 10^{-26}\left(\frac{B_{\rm 0,\perp}}{10^{-9}\,{\rm G}}\right)
\left(\frac{M_{\rm a}}{10^{11}\,{\rm GeV}}\right)^{-1}{\rm cm}^{-1},\nonumber \\
\Delta_{\rm m} & = & -2.5 \cdot 10^{-28}
\left(\frac{m_{\rm a}}{10^{-16}\,{\rm eV}}\right)^2
\left(\frac{\omega}{1\,{\rm eV}}\right)^{-1}{\rm cm}^{-1},
\end{eqnarray}
where $\omega$ is the photon energy, $n_{\rm e}$ is the electron
density, $B_{0,\perp}$ is the strength of the magnetic field
perpendicular to the propagation of the photon, $M_{\rm a}$ is the
inverse coupling between the photon and the axion, and $m_{\rm a}$ is
the mass of the axion. The initial condition of the equation is
\begin{equation}
  \label{eq:rho0} 
  \rho_{\rm 0}=\left(\begin{array}{ccc}
       \frac{1}{2} & 0 & 0\\
       0 & \frac{1}{2} & 0\\
       0 & 0 & 0
       \end{array}\right),
\end{equation}
corresponding to unpolarised light. The diagonal elements correspond
to the intensities of the two polarisation states of the photon and
the intensity of the axion, respectively. The system of 9 coupled
(complex) differential equations is solved numerically using the
SuperNova Observation Calculator (SNOC) \cite{Goobar:2002vm}. Assuming
a quasar (or a SN) at a certain redshift, the light beam from the
source is followed through a large number of spherical cells. Each
cell has a value for the magnetic field and the electron density. The
electron density and the magnetic field strength are evolved as $n_e
\propto (1+z)^3$ and $B \propto (1+z)^2$. The dispersion of the field
strength and the density is set to 50\% of their values. We assume
that the magnetic field is frozen into the plasma and subsequently
given by $B_{\rm 0}\propto n_{\rm e}^{2/3}$ with random direction
\cite{Christensson:2002ig}. We also test whether our simulations are
in accordance with current Faraday rotation measurements. Since the
magnetic domains are small compared to the total travel length and the
magnetic fields have random directions, the induced Faraday rotation
is too small to be used to rule out any of our investigated models.
As the beam pass through the cells, the background cosmology and the
wavelength of the photon are updated, as are the matrices $\rho$ and
$M$. In all of our simulations, we use $[\om =0.3, \ola =0.7,
h=0.7]$. However, the cosmology dependence is weak.

\subsection{Parameter dependence}
\label{sec:param}

The polarisation state of the photon is disregarded by rewriting $M$
as a $2 \times 2$ matrix,
\begin{equation}
  \label{eq:m2D} 
  M^{\rm 2D}=\left(\begin{array}{ccc}
       \Delta & \Delta_{\rm M}\\
       \Delta_{\rm M} & \Delta_{\rm m}
       \end{array}\right),
\end{equation}
where $\Delta = \Delta_{\perp} = \Delta_{\parallel}$. After solving
equation (\ref{eq:ih}) analytically we find that the photon intensity, $\rho^{\rm
2D}_{11}$, is
\begin{eqnarray}
  \label{eq:rho11} 
  \rho^{\rm 2D}_{11}&=&1-\frac{\Delta_{\rm M}^2}{2\Omega^2}
  (1-\cos{\Omega t}),\nonumber \\
  \Omega &=&\frac{1}{2}\sqrt{(\Delta -\Delta_{\rm m})^2+4\Delta_{\rm
  M}^2}.
\end{eqnarray}

Using the quantities in equation (\ref{eq:terms}) we see that $|\Delta_m|
\gg |\Delta|$ when $m_{\rm a} \gg 12 \sqrt{10} \sqrt{n_e/(10^{-8}~{\rm
cm}^{-3})} 10^{-16}~{\rm eV} \approx 38\sqrt{n_e/(10^{-8}~{\rm
cm}^{-3})} 10^{-16}~{\rm eV} $. In our parameter interval for $n_e$,
$B_0/M_{\rm a}$, and $\omega$, we then have $|\Delta_m| \gg
|\Delta_M|$ and $\Omega \simeq \frac{1}{2}|\Delta_m|
\propto m_{\rm a}^2\omega^{-1}$. For $m_{\rm a} \gg 38\sqrt{n_e/(10^{-8}~{\rm
cm}^{-3})} 10^{-16}~{\rm eV}$, oscillations are thus suppressed as
$m_{\rm a}^{-4}$ and the dimming is negligible.

From now on, we assume that $m_{\rm a} \ll 38\sqrt{n_e/(10^{-8}~{\rm
cm}^{-3})}
 10^{-16}~{\rm eV}$ where the oscillation probability is
independent of
 the axion mass and the parameters governing the
oscillation
 probability are the plasma density, $n_{e0}$, and the
magnetic field
 divided by the inverse coupling between the photon and
the axion,
 $B_0/M_{\rm a}^{11}$ \footnote{In the future we will use
the notation
 $M_{\rm a}^{11} \equiv M_{\rm a}/ (10^{11}~{\rm
GeV}).$}, at $z=0$. In this case we have $|\Delta_m| \ll |\Delta|$ and
thus $\Omega \simeq
\frac{1}{2}\sqrt{\Delta^2+4\Delta_{\rm M}^2}$. 
For $B\omega/(M_{\rm a}n_e) \gg 9 \cdot 10^{-21}~{\rm cm^3G}$, we have $|\Delta|
\ll 2|\Delta_{\rm M}|$ and $\Omega \simeq \Delta_{\rm M}$, 
i.e., the oscillations are independent of energy. This is the most
important difference from the results obtained in
reference \cite{Mortsell:2003ya}, making it impossible to exclude any
regions to the left of $B\omega/(M_{\rm a}n_e) \approx 9 \cdot
10^{-21}~{\rm cm^3G}$ (see, e.g., figure~\ref{fig:contour}).

\subsection{Method}

An effect of photon-axion oscillations is that a dispersion is added
to the quasar spectra due to the energy dependence of the effect. By
comparing the dispersion in observed quasar spectra with the
dispersion in simulated quasar spectra, we can conclude if the model
behind each simulation is allowed.

\subsection{Observed quasar spectra}
\label{sec:realqs}

\begin{figure}
  \begin{center}
    \epsfxsize = 10cm
    \epsffile{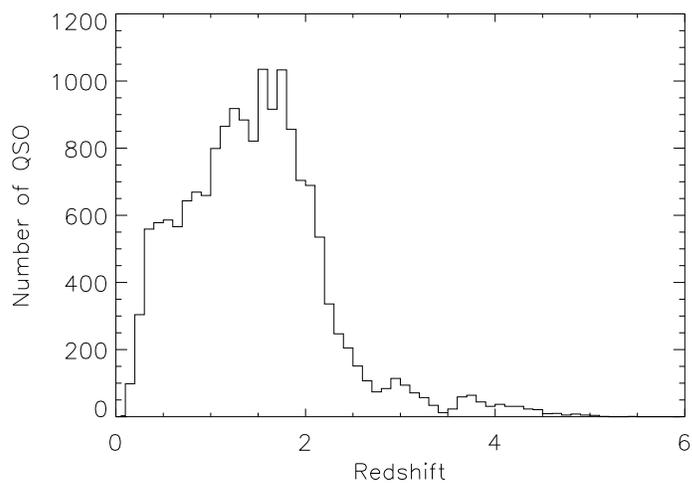}

    \caption{Redshift distribution of the quasars in the SDSS DR1. The
    redshift bins in the figure have a width of $\Delta z = 0.1$.}

    \label{fig:zdistr}
  \end{center}
\end{figure}

The spectra of 16\,713 quasars are obtained from the SDSS DR1
\cite{Schneider:2003zt}. These quasars have redshifts between
$z=0.08$ and $z=5.41$ with a median redshift of $z=1.43$. The redshift
distribution is shown in figure~\ref{fig:zdistr}.

The spectra are divided into redshift bins of size $\Delta z = 0.2$
where we only consider the redshift interval $0.1<z<2.9$ since we need
a certain number of quasars in each bin to obtain acceptable
statistics. In this redshift range we have 15\,914 quasars.

\begin{figure}
  \begin{center}
    \epsfxsize = 12cm
    \epsffile{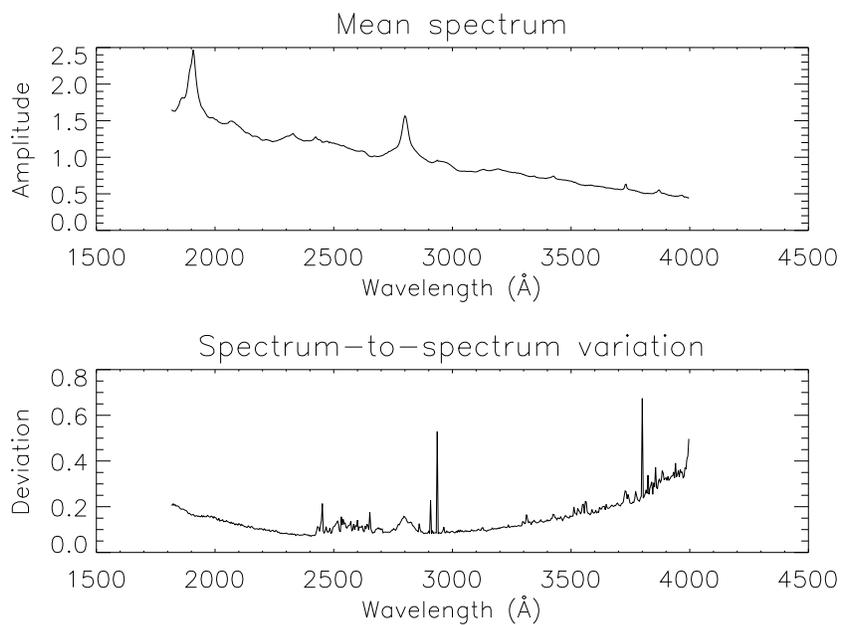}

    \caption{The mean rest frame spectrum and the spectrum-to-spectrum
    variation for the SDSS quasars with redshifts in the interval
    $1.1<z<1.3$.}

    \label{fig:sigspec}
  \end{center}
\end{figure}

Each spectrum is redshifted into its restframe and divided into
wavelength bins of size $4$ {\AA} and the average flux is calculated
in each of these bins for the individual spectra. We calculate the
mean spectrum along with the dispersion of the individual spectra from
the mean spectrum in each redshift bin. All spectra more than three
standard deviations from the mean are rejected in order to minimize
the effect from peculiar spectra. An example of a mean spectrum and
the spectrum-to-spectrum variation for a particular redshift bin is
given in figure~\ref{fig:sigspec}.

\subsection{Simulated quasar spectra}

\begin{figure} [t]
  \begin{center}
    \epsfxsize = 12cm
    \epsffile{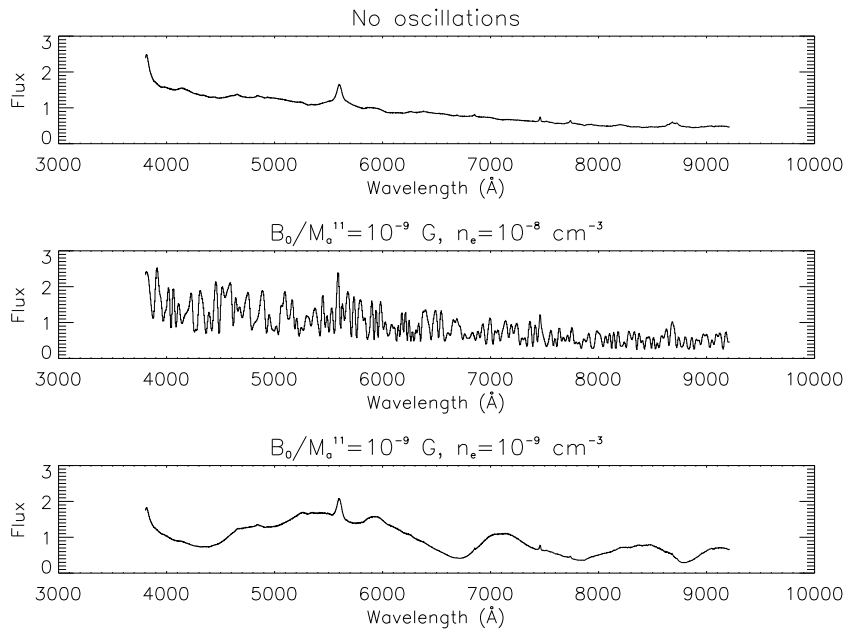}
    \caption{Simulated mean quasar spectra for three different
    scenarios when $z=1$.}
    \label{fig:oscfig}
  \end{center}
\end{figure}

Using SNOC, we simulate the effects of photon-axion oscillations on
quasar observations for $10^{-12}~ {\rm
cm^{-3}}<n_{e0}<3\cdot10^{-7}~{\rm cm^{-3}}$ and $10^{-11}~{\rm
G}<B_0/M_{\rm a}^{11}<10^{-8}~{\rm G}$, using the median quasar
spectrum obtained in reference \cite{VandenBerk:2001hc}. Note that the
upper limit of the electron density is motivated by the WMAP
measurements of the baryon density which gives a maximum allowed
plasma density of $n_e = 2.7\cdot 10^{-7} {\rm
cm^{-3}}$\cite{Spergel:2003cb}. In figure~\ref{fig:oscfig} we have
plotted the normalised quasar spectrum for three different oscillation
scenarios for $z=1$.

For each redshift bin and set of parameter values, a large number of
quasar spectra are generated and the mean spectrum and the dispersion
due to the oscillations is calculated.  Comparing the dispersion in
the simulated and the observed quasar spectra, we are then able to
constrain the values of $n_{e0}$ and $B_0/M_{\rm a}^{11}$. If the
simulated dispersion is smaller than the observed, we cannot exclude
the scenario since real quasars have an intrinsic dispersion. If the
dispersion is larger for the simulated spectra, a comparison is made
between the simulated and the observed cumulative distribution of
dispersions using the Kolmogorov-Smirnov test. The test gives the
significance level that the distributions differ significantly for the
particular redshift bin. The total significance level for a certain
set of parameter values is calculated as the product of all
significance levels for the different redshift bins.

For each scenario we also calculate the rest frame B-band attenuation,
due to the photon-axion oscillations for a SNIa at $z=1$.

\subsection{Result}

\begin{figure} [t]
  \begin{center}
    \epsfxsize = 12cm
    \epsffile{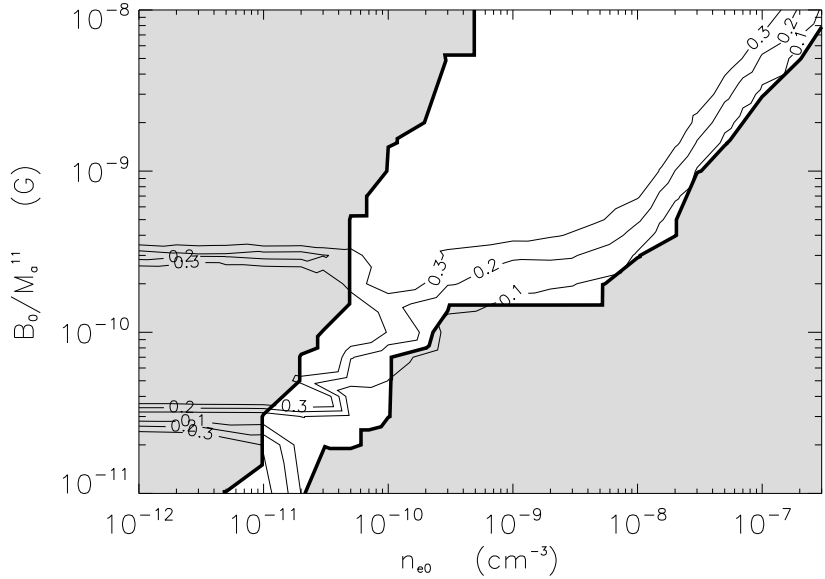}
    \caption{The grey region indicates the allowed parameter region
    at
 a confidence level of 95\%. The lines labeled [0.3, 0.2, 0.1]
    show the B-band attenuation expressed in magnitudes for a SNIa at
    $z=1$.}
    \label{fig:contour}
  \end{center}
\end{figure}

\begin{figure}
  \begin{center}
    \epsfxsize = 12cm
    \epsffile{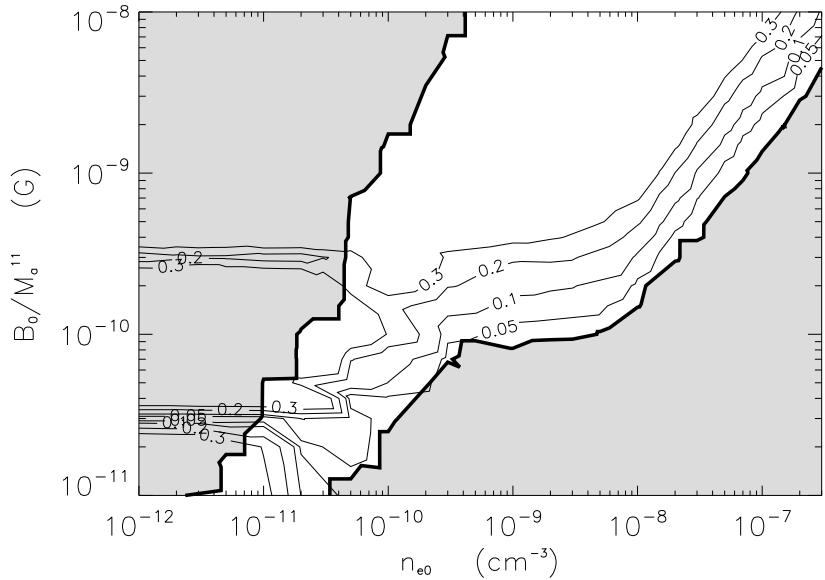}

    \caption{The grey region indicates the allowed parameter region at
	a confidence level of 95\% when an intrinsic dispersion of 5\% has
	been added to the dispersion of the simulated spectra. The thin lines
	show the B-band attenuation expressed in magnitudes for a
    	SNIa at $z=1$.}

    \label{fig:qsointr}
  \end{center}
\end{figure}

In figure~\ref{fig:contour}, it is shown what values of the electron
density and the magnetic field strength are allowed if there exists an
axion with a mass $m_{\rm a} < 10^{-16}~{\rm eV}$. The figure also
contains the corresponding attenuation for a SNIa.
The small scale features of the plot is an artefact of the choice of
simulated values. If $B\omega/(M_{\rm a}n_e) \lesssim 9
\cdot 10^{-21}{\rm ~cm^3G}$, the dimming of SNIa at $z=1$ due to
oscillations is less than 0.1 magnitudes. If $B\omega/(M_{\rm a}n_e)
\gtrsim 9\cdot 10^{-21}{\rm cm^3G}$, the oscillation probability is
independent of energy and a dimming as large as 0.56 magnitudes for a
SNIa at $z=1$ cannot be ruled out. Employing the CAST lower limit on
$M_{\rm a}$ \cite{Andriamonje:2004}, the oscillation probability is
independent of energy for $B\omega/n_e \gtrsim 8\cdot
10^{-2}~{\rm cm^3 eV G}$. Note that our
conclusions are conservative in the respect that we have not added any
intrinsic dispersion to the simulated quasar spectra. Assuming an
intrinsic dispersion of 5\%, the upper limit on the dimming of SNIa at
$z=1$ becomes less than 0.05 magnitudes when $B\omega/(M_{\rm a}n_e)
\lesssim 9 \cdot 10^{-21}{\rm ~cm^3G}$ (see figure~\ref{fig:qsointr}).

\subsection{Comparison with the earlier analysis}

The larger sample of the DR1 compared to the EDR leads only to minor
changes in the conclusions. However, the analysis of the effects of
photon-axion oscillations using EDR data in M\"{o}rtsell and Goobar
\cite{Mortsell:2003ya} contained an error in the evolution of the
density matrix (corresponding to equation (\ref{eq:ih}) in this
paper). Instead of $2{\rm i}\frac{\partial \rho}{\partial t} = [{\cal
M},\rho]$, they used $2\omega{\rm i} \frac{\partial \rho}{\partial t}
= [{\cal M},\rho]$. Disregarding the polarisation state of the photons
(see Section \ref{sec:param}), the photon intensity, $\rho^{\rm
2D}_{11}$, would then be given by\footnote{The corresponding equation
in reference
\cite{Mortsell:2003ya} had also a factor of two wrong due to a
misprint. This factor was not used in the calculations.}
\begin{eqnarray}
  \rho^{\rm 2D}_{11}&=&1-\frac{\Delta_{\rm M}^2}{2\omega^2\Omega^2}
  (1-\cos{\Omega t}),\nonumber \\
  \Omega &=&\frac{1}{2\omega}\sqrt{(\Delta -\Delta_{\rm m})^2+4\Delta_{\rm
  M}^2}.
\end{eqnarray}
Because of the $\omega$ in the denominator of the expression for
$\Omega$, the oscillation probability never became independent of
wavelength. Therefore, the upper left region of the plot could be excluded.

Another difference compared to the earlier analysis is that
$m_a=10^{-16}~{\rm eV}$ was used in \cite{Mortsell:2003ya}, while $m_a
<< 10^{-16}~{\rm eV}$ has been used in this paper. This only affects
the region where both $n_{e0}$ and $B_{0}/M_a^{11}$ are small.

\subsection{Dimming evolution}

\begin{figure}
  \begin{center}
    \epsfxsize = 10cm
    \epsffile{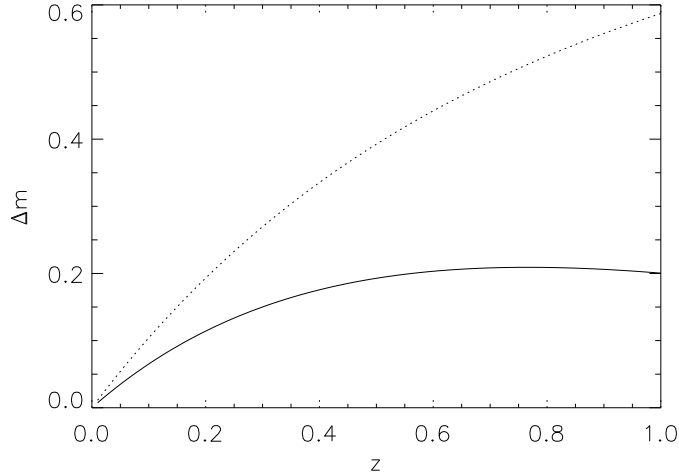}

    \caption{The difference in magnitude, $\Delta
    m(\om,\ola)$, as a function of redshift for
    different values of $\om$ and $\ola$. The solid
    line corresponds to $m(0.3,0.7)-m(0.3,0)$ and the dotted line to
    $m(0.3,0.7)-m(1,0)$.}

    \label{fig:dimdiff}
  \end{center}
\end{figure}

\begin{figure}
  \begin{center}
    \epsfxsize = 12cm
    \epsffile{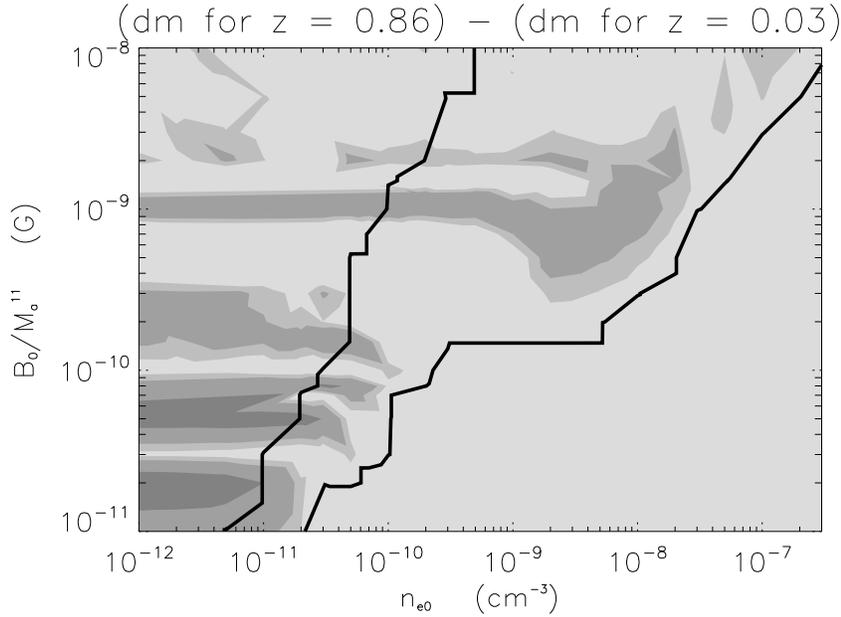}

    \caption{The solid line indicates the border to the allowed
    parameter region. The different shades indicate regions with
    seperate differences in attenuation between a SNIa at $z=0.86$ and
    one at $z=0.03$. The borders to these regions correspond to 0.15,
    0.25 and 0.50 magnitudes. The palest region indicates a difference
    in attenuation of less than 0.15 magnitudes.}

    \label{fig:contour_diff}
  \end{center}
\end{figure}

\begin{figure}
  \begin{center}
    \epsfxsize = 10cm
    \epsffile{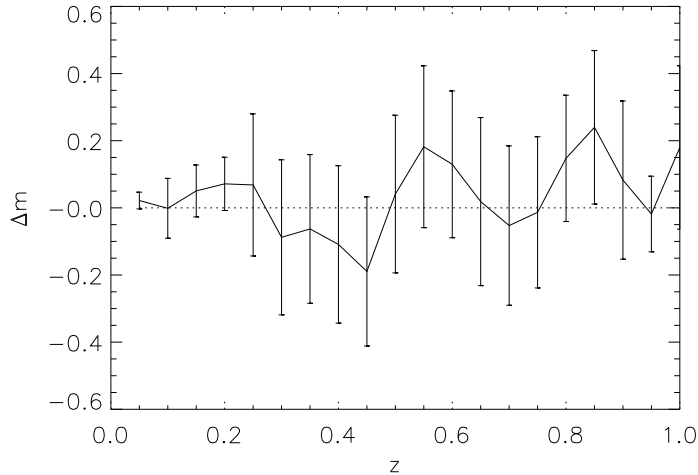}

    \caption{The difference in magnitude, $\Delta m$, as a function of
    redshift for $(\om,\ola)=(0.3,0.7)$ (no oscillations) and an open
    universe with $(\om,\ola)=(0.3,0)$, $n_e = 5\cdot10^{-12} {\rm
    cm^3}$ and $B_0/M_{\rm a}^{11} = 2 \cdot 10^{-10} {\rm G}$. The
    attenuation is non-monotonic since the oscillation length is
    smaller than the distance to $z\sim 1$. The error bars correspond
    to the dispersion induced by inhomogeneities in the magnetic field
    strength and electron density (see also Section \ref{sec:disp}).}

    \label{fig:bzdep_diff}
  \end{center}
\end{figure}

If the attenuation caused by photon-axion oscillations is large both
at low and high redshifts, this cannot explain the faintness of
distant SNe, even though it is still possible that the oscillations
exist and lead to a dimming of all SNe. Thus, we need to investigate
the axion attenuation as a function of redshift. In
figure~\ref{fig:dimdiff}, the difference in observed magnitude for
different cosmologies is plotted.
If the universe has $(\om,\ola) = (0.3,0)$ and the observed SN
magnitudes is best fitted by a cosmology with $(\om,\ola)=(0.3,0.7)$,
oscillations need to account for a difference in dimming of 0.20
magnitudes between a SNIa at $z=0.86$ and one at $z=0.03$. If instead
the universe has $(\om,\ola) = (1,0)$ we need to be able to explain a
dimming of 0.52 magnitudes between $z=0.86$ and $z=0.03$. In
figure~\ref{fig:contour_diff}, we have plotted the difference in
attenuation between $z=0.86$ and $z=0.03$ for different values of the
parameters $n_e$ and $B_0/M_{\rm a}^{11}$\footnote{Note that the
dimming of the simulations are calculated for a universe where
$(\om,\ola) = (0.3,0.7)$. The effect of the cosmology is, however,
small and can be effectively cancelled by small changes in the
magnetic field strength and the electron density.}.
As an example, a universe with $(\om,\ola)=(0.3,0)$, $n_e=5 \cdot
10^{-12} {\rm cm^3}$ and $B_0/M_{\rm a}^{11} = 2 \cdot 10^{-10} {\rm
G}$ where photons mix with axions with a mass $m_{\rm a} \lesssim
10^{-16}~{\rm eV}$ could yield almost exactly the same observed
magnitudes as a cosmology with $(\om,\ola)=(0.3,0.7)$ without
oscillations (see figure \ref{fig:bzdep_diff}). This scenario is not
possible to rule out using the observed (lack of) dispersion in
quasar
 spectra. Note however that in this specific case, the
oscillation
 length is smaller than the distance to $z\sim 1$. In
order to get an
 attenuation that grows monotonically out to $z\sim
1$, we need an
 oscillation length of the order the Hubble length. For
the region where
 the effect is energy independent, this corresponds
to $B_0/M_{\rm
 a}^{11}\sim 10^{-12} - 10^{-11}{\rm G}$. We stress
again that
 since cosmic microwave background observations show that
the universe
 is close to flat and large scale galaxy surveys indicate
a low matter
 density, it is unlikely that all the dimming is caused
by photon-axion
 oscillations. However, as for today, it is not
possible to rule out
 substantial systematic effects from photon-axion
oscillations.

\subsection{Dispersion in type Ia supernova magnitudes}
\label{sec:disp}

Since the oscillation probability is independent of photon energy for
$B\omega/(M_{\rm a}n_e) \gtrsim 9 \cdot 10^{-21}{\rm ~cm^3G}$ we
cannot use the induced dispersion in quasar spectra to constrain the
effect in this region. As an additional constraint, we can instead use
the induced dispersion in SNIa magnitudes. If the magnetic field
strength and electron density is not perfectly homogeneous, different
lines of sight will correspond to different phases in the photon-axion
oscillation. This will cause a dispersion also in integrated broad
band magnitudes of high redshift sources (see
figure~\ref{fig:bzdep_diff}). Since the dispersion can be of the same
order as the attenuation, we should be able to constrain the effect by
studying the observed dispersion in SNIa magnitudes.

\begin{figure}
  \begin{center}
    \epsfxsize = 10cm
    \epsffile{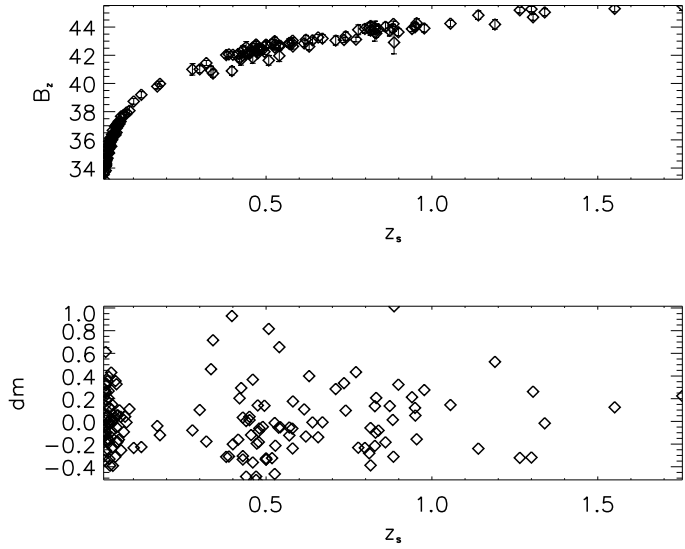}

    \caption{The B magnitude and the dispersion in the B magnitude as
    a function of redshift for some of the SNe in
    reference \cite{Riess:2004nr}.}

    \label{fig:sn}
  \end{center}
\end{figure}

\begin{figure}
  \begin{center}
    \epsfxsize = 10cm
    \epsffile{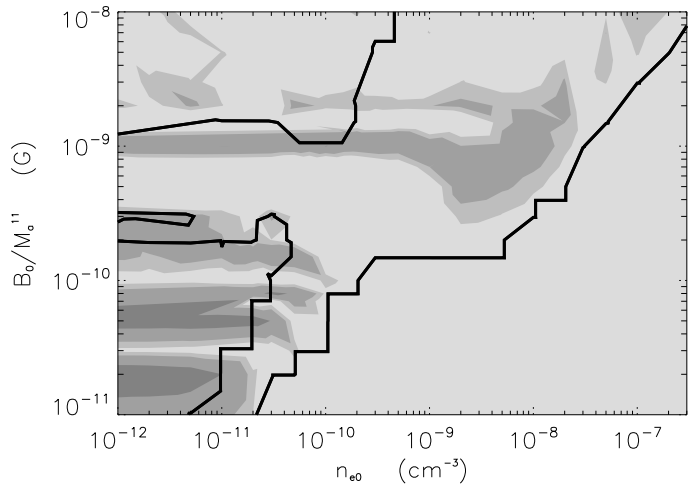}

    \caption{The solid line indicates the border to the allowed
    parameter region combining current quasar constraints and future
    constraints using the dispersion in SNIa magnitudes. An intrinsic
    dispersion of 0.1 magnitudes for 50 SNe in each redshift bin
    $z=0.03, 0.49$ and $0.86$ is assumed. The palest region indicates
    a difference in attenuation of less than $0.15$ magnitudes between
    a
 SNIa at $z=0.86$ and one at $z=0.03$. The darker regions
    indicates a difference of $0.15-0.25$, $0.25-0.50$ and $> 0.50$ magnitudes.}

    \label{fig:contour_disp}
  \end{center}
\end{figure}

In figure~\ref{fig:sn}, we plot the B magnitudes of the SNIa in
reference \cite{Riess:2004nr} as a function of redshift.
We construct three redshift bins ($0<z<0.1$, $0.3<z<0.7$, and
$0.7<z<1.1$) with the mean redshifts 0.03, 0.49, and 0.86,
respectively. The magnitude dispersion in the bins are $0.22 \pm
0.02$, $0.33 \pm 0.04$, and $0.30 \pm 0.06$. We use the first
redshift bin to estimate the intrinsic dispersion of the SN
magnitudes as $\sigma_{\rm intr} = \sqrt{\sigma_{\rm real}^2 -
\sigma_{\rm sim}^2}$, where $\sigma_{\rm real}$ is the dispersion of
the observed SNe and $\sigma_{\rm sim}$ is the simulated dispersion
caused by photon-axion oscillations at this redshift. We can then
correct the simulated dispersions in the higher redshift bins for the
intrinsic dispersion and check if the dispersion is larger than
observed. For current data, the dispersion is too large to give any
useful constraints. If we in the future have a larger SN sample and a
smaller dispersion we would be able to diminish the allowed parameter
region, but we would not be able to improve the limits on the dimming,
see figure~\ref{fig:contour_disp}.

\section{Dust}
\label{sec:dust}

Dust grains are present both in the interplanetary, interstellar, and
intracluster space. It is not known with certainty if there is dust in
the intergalactic space. Dust in the Milky Way can be described by a
mean extinction law $A(\lambda)= f(R_V)$ where $A(\lambda)$ is the
extinction for the wavelength $\lambda$. Reddening is a consequence of
the fact that shorter wavelengths are more effectively absorbed and
scattered by dust than longer. It is quantisised by the reddening
parameter, $R_V$, which is defined as
\begin{equation}
R_V \equiv \frac{A_V}{E(B-V)},
\end{equation}
where $A_V$ is the V-band extinction and $E(B-V)$ is the colour excess,
\begin{equation}
E(B-V) \equiv (B-V)_{\rm obs}-(B-V)_{\rm intr}.
\end{equation}
Here $(B-V)_{\rm obs}$ is the observed colour and $(B-V)_{\rm intr}$
is the intrinsic colour. Within the Milky Way, we have $2\lesssim R_V
\lesssim 6$, but $R_V\approx 3.1$ is often a useful approximation. 
Among other galaxies, values of $R_V$ has been observed in the
interval $1.5 \lesssim R_V\lesssim 7.2$ \cite{Falco:1999jc}. Since
small particles preferentially scatter light with short wavelengths, a
population of small grains have a lower value of $R_V$ than a
population of large grains. Different parametrisations of $f(R_V)$ is
given in, e.g., Cardelli, Clayton and Mathis (CCM)
\cite{Cardelli:1989fp} and Fitzpatrick \cite{Fitzpatrick:1999}. 
In reference~\cite{Mortsell:2003vk}, the CCM parametrisation was
used. In this paper, we compare our results using the different forms
of $f(R_V)$, finding that the differences are negligible except for
very high values of $R_V$. Note however that it is not known if any of
the parametrisations are valid for an intergalactic dust population
with $R_V\gtrsim 7$. The results presented in this work are obtained
using the $f(R_V)$ given in Fitzpatrick \cite{Fitzpatrick:1999}.

\subsection{Constraining intergalactic dust}
\label{sec:dustconstraint}

The main constraints on the intergalactic dust population are:
\begin{enumerate}

	\item The dust population must not produce such an amount of
	reddening that it causes discrepancies with observations.

	\item The dust must not contribute with more radiation to the
	background, than what is observed.

\end{enumerate}
The intergalactic dust absorbs energy from the optical and the UV
background and then emits it in the far-infrared. This can be used to
constrain the intergalactic dust since it cannot contribute with more
radiation in this wavelength region than what has not been identified
as other sources. A universe where the dimming of SNIa is explained
solely by intergalactic dust probably cannot have $(\om,\ola) =
(1,0)$, while $(\om,\ola) = (0.2,0)$ cannot be ruled out
\cite{Aguirre:1999uu}.

In this paper we will use the reddening of high redshift sources to
constrain the properties of intergalactic dust and the effects on SN
observations (see also reference \cite{2002A&A...384....1G}). This is
accomplished with colour comparisons of real quasars with simulated
quasars for different dust scenarios.

\subsection{Theory of the simulations}
\label{sec:theory}

We assume a quasar (or a SN) at a certain redshift and follow the
photons emitted from the source to the detector at Earth using SNOC
\cite{Goobar:2002vm}. The path of the photons is divided into cells
of the same size as the average distance between galaxies. In each
cell, the dust density and the differential extinction coefficient is
specified and the extinction due to intergalactic dust is calculated.

The dust attenuation, $\Delta m_{\rm dust}$, at a given emission
redshift $z_e$ and a given observed wavelength is given by
$\lambda_o$ \cite{Mortsell:2003vk}
\begin{equation}
\label{eq:deltam}
\Delta m_{\rm dust} (z_e,\lambda_o) = \frac{2.5}{\ln{10}} \int_{0}^{z_e} 
{\frac{A(\lambda_o/(1+z),R_V)/A(V)}{D_V(z)H(z)(1+z)} ~dz },
\end{equation}
where $A(V)$ is the attenuation in the V-band and $D_{V}$ is the
interaction length for photon scattering or absorption with dust
particles in the V-band.  The function $H(z)$ is given by
\begin{equation}
    H(z)= H_0 \sqrt{(1+z)^3\,\om +(1+z)^2\,(1-\om-\ola)+\ola}\, .
\end{equation}

We study two different models of the dust distribution with $\rho_{\rm
dust} \propto (1+z)^{\alpha(z)}$ and $\alpha(z)$ given by
\cite{Goobar:2002vm,Mortsell:2003vk}:

\begin{tabular}{l l}
Model A: & $ \alpha(z) = 3~\forall~z$ \\ Model B: & $ \alpha(z) =
\left\{ \begin{array}{ll} 3~\forall~z\le0.5 & \\ 0~\forall~z>0.5 & \\
\end{array} \right.$ \\
\end{tabular} \\
Model A corresponds to a dust population with
a constant comoving density whereas model B corresponds to a
population where dust is created at redshifts $z>0.5$ with the same
rate as the dilution caused by the expanding universe. The two models
represent two extremes in a scale of dust models and they can
therefore be considered to capture a wide range of possible dust
models. 

Since $D_{V}(z) = [\sigma n_{\rm dust}(z)]^{-1}$, where $\sigma$ is
the interaction cross section and $n_{\rm dust}(z)$ is the dust number
density, we have $D_{V}(z) \propto \rho_{\rm dust}^{-1}\propto (1+z)^{
-\alpha(z)}$. The proportionality constant is determined by $D_{0V}$,
the interaction length for the V-band at zero redshift.

\subsection{Observed quasar colours}
\label{sec:sample} 

We use the same sample of 16\,713 quasars from the SDSS DR1
\cite{Schneider:2003zt} described in Section \ref{sec:realqs}. 
The 821 quasars that have been classified as extended are excluded to
avoid problems with contamination from the host galaxy.

To avoid quasars with too much noise, we only include objects with
magnitudes brighter than 22.3, 22.6, 22.7, 22.4, and 20.5,
respectively for the five filters (u, g, r, i, and z), corresponding
to a signal-to-noise (S/N) ratio better than 5:1
\cite{Richards:2000et}. This leaves 15\,194 quasars.

We also restrict ourselves to redshifts $0.5<z<2$. This is primarily
because we want to have good statistics, i.e., many quasars in every
redshift bin. Another reason for excluding objects with redshifts
$z<0.5$ is that quasar spectra at these redshifts often are
contaminated by their host galaxy. Finally we have 11\,694 point-like
quasars with redshifts $0.5<z<2$ that fulfill the S/N limit.

The quasars are divided into redshift bins of size $\Delta z =
0.05$. For each bin, the mean colour, $(X-Y)_z^{\rm obs}$, is
calculated for all combinations of different filters, X and Y.
Objects with a colour more than two standard deviations from the mean
value is rejected. This is because we are not interested in using
peculiar quasars that have evolved under unusual circumstances.

\subsection{Simulated quasar colours}

We assume a universe with $[\om =0.3, \ola =0.7, h=0.7]$ and simulate
dust scenarios with $10~{\rm Gpc} \le D_{0V} \le 5000~{\rm Gpc}$ and
$0 \le R_V \le 12$. Each parameter set ($R_V$,$D_{0V}$) is simulated
twice, once for each dust model described in Section
\ref{sec:theory}. For each scenario, a large number of quasar events
are simulated with different values of the redshift. For each event,
the K correction for the SDSS broadband filters u, g, r, i, and z is
generated using the median quasar spectrum obtained in reference
\cite{VandenBerk:2001hc}. The simulated mean colour,
$(X-Y)_{z,R_V,D_{0V}}^{\rm sim}$, is calculated by subtracting the K
corrections for the filters, which is the same as subtracting the two
filter magnitudes.

The probability of the different scenarios are determined with the
$\chi^2$-function
\begin{eqnarray}
\chi^2 = \sum_{i,j=1}^N \Delta(X-Y)_{z_i,R_V,D_{0V}}(V(X-Y)^{-1})_{i,j}
\Delta(X-Y)_{z_j,R_V,D_{0V}},&\\
\Delta(X-Y)_{z,R_V,D_{0V}} = (X-Y)_z^{\rm obs}-(X-Y)_{z,R_V,D_{0V}}^{\rm sim},&
\end{eqnarray}
where $i$ and $j$ refer to the redshift bins. The covariance matrix
$V(X-Y)_{i,j}$ is defined as
\begin{equation}
V(X-Y)_{i,j} = \sigma(X)^2_{\rm sys}+\sigma(Y)^2_{\rm sys}+ \delta_{ij}
\left[\sigma(X)^2_{\rm tem}+\sigma(Y)^2_{\rm
tem}+\sigma(X-Y)_{\rm obs}^2\right],
\end{equation}
where $\sigma_{\rm sys}(i)$ is the systematic error and $\sigma_{\rm
tem}(i)$ is the template error associated with the broadband filter
$i$. The template errors arise because we use a median quasar spectrum
in the simulation of the K corrections and not all rest frame quasar
spectra are identical. The systematic filter errors are 0.03 for the u
and z filters and 0.02 for the others, while the template errors have
been estimated to 0.1 for the u and z filter and to 0.05 for the other
filters. $\sigma_{\rm obs}$ denotes the deviation from the mean
observed colour in each redshift bin. Note that the error budget is
completely dominated by the template error. The inclusion of quasar
data from the second and third data release of the SDSS will
therefore have a small effect on the constraints obtained.

\subsection{Result}

\begin{figure}
  \begin{center} \epsfxsize = 12cm \epsffile{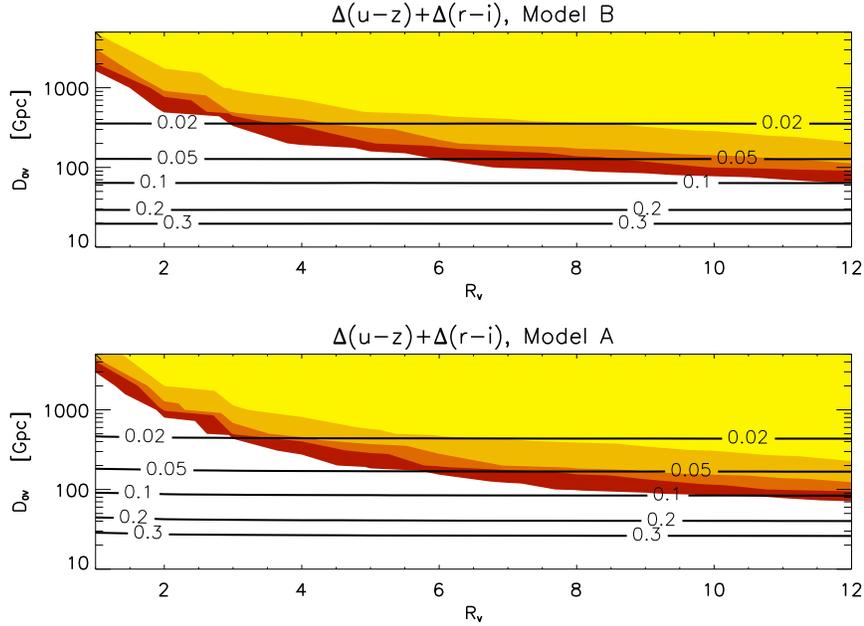}

    \caption{The confidence levels combining $\Delta(u-z)$ and
    $\Delta(r-i)$ for the two dust models and the attenuation of a
    SNIa at $z=1$. The palest region indicates the 68\% confidence
    level allowed by the $\chi^2$-test and the darker regions indicate
    90\%, 95\%, and 99\%. The almost horizontal lines show the B-band
    attenuation in magnitudes caused by dust for a SNIa at $z=1$.}

    \label{fig:dust_ftz_uzri}
  \end{center}
\end{figure}

In figure~\ref{fig:dust_ftz_uzri}, we present our results from the
$\chi^2$-analysis combining $\Delta(u-z)$ and $\Delta(r-i)$ together with
the expected dimming in the B-band, $\Delta m_B(z)$, for a SNIa at the
$z=1$. We conclude that at 99\% confidence level, the dimming in the
B-band $\Delta m_B\lesssim 0.1$ for a SNIa at $z=1$. We can also limit
the interaction length for the V-band at $z=0$ to be larger than
$60~{\rm Gpc}$. If the intergalactic dust is similar to the Milky Way
dust (with $R_V \lesssim 4$), we are able to rule out $\Delta m_B
\gtrsim 0.03$.

If we use the CCM parametrisation instead of the parametrisation by
Fitzpatrick we get an upper limit on the B-band dimming of a SNIa at
$z=1$ of 0.2 magnitudes.

Note that the scenario without any intergalactic dust is not
eliminated by the analysis. Also, the increase in the number of
quasars and the improvement of the systematic filter errors of the
SDSS DR1 compared to the EDR have led only to minor improvements
compared to the results obtained in reference \cite{Mortsell:2003vk}.

\section{Summary}
\label{sec:summary}

Observations of high redshift SNIa show that the universe is dominated
by dark energy accelerating the expansion rate. In order to take
advantage of the increased SN statistics and to be able to use SNIa
data to determine the dark energy properties, it is necessary to
control the systematic effects involved. We have used quasar colours
and spectra from the SDSS DR1 to constrain the effects from
intergalactic dust attenuation and photon-axion oscillations,
respectively.

We have found that the largest possible dimming of a SNIa at $z=1$
caused by intergalactic dust is 0.2 magnitudes. The addition of new
quasars to the intergalactic dust analysis will not improve the limits
since the analysis is limited by systematic errors, in particular the
template errors. One way of improving the template errors is to use
different template spectra for different groups of quasars. For
example, a median spectrum created from low redshift quasars will be
unaffected by possible intergalactic dust and photon-axion
oscillations. If the template errors were half their size, we would be
able to give a maximum dimming limit of 0.05 magnitudes for a SNIa at
$z=1$. Also, improving the resolution of the FIRB experiments could
lead to the rejection of more dust scenarios.  The calculations of the
dimming caused by intergalactic dust is based on mean extinction laws
for the dust in the Milky Way
\cite{Cardelli:1989fp,Fitzpatrick:1999}. It is not known if these
parametrisations are valid for an intergalactic dust population where
$R_V$ might be much larger than in the interstellar medium. Also, we
have assumed that the reddening parameter, $R_V$, is independent of
direction and redshift. However, this is a fair approximation since
the simulated SN attenuation is almost independent of $R_V$ (see
e.g. figure~\ref{fig:dust_ftz_uzri}).

For photon-axion oscillations we have found a maximum dimming of 0.1
magnitudes for SNIa at $z=1$ if $B\omega/(M_{\rm a}n_e) \lesssim 9
\cdot 10^{-21}{\rm ~cm^3G}$, where $B$ is the magnetic field strength,
$\omega$ is the photon energy, $M_{\rm a}$ is the inverse coupling
between the photon and the axion, and $n_e$ is the electron
density. However, when $B\omega/(M_{\rm a}n_e) \gtrsim 9 \cdot
10^{-21}{\rm ~cm^3G}$ (corresponding to $B\omega/n_e \gtrsim 8\cdot
10^{-2}~{\rm eVcm^3G}$ using the CAST lower limit on $M_{\rm a}$
\cite{Andriamonje:2004}), the oscillation
 probability is independent
of photon energy and a dimming as large as
 0.6 magnitudes cannot be
excluded. We have also shown that for certain
 combinations of $n_{e}$
and $B/M_{\rm a}^{11}$, specifically
 $B\omega/(M_{\rm a}n_e) \gtrsim
9 \cdot 10^{-21}{\rm ~cm^3G}$ and
 $B_0/M_{\rm a}^{11}\sim 10^{-12} -
10^{-11}{\rm G}$, the attenuation
 from oscillations evolve with
redshift in a manner very similar to the
 observed SNIa dimming. The
region where the effect is independent of
 photon energy could be
constrained using the observed dispersion in
 SNIa magnitudes. An
inhomogeneous magnetic field strength and/or
 electron density will
cause different lines of sight to correspond to
 different phases in
the photon-axion oscillation. Thus we expect a
 dispersion in
integrated broad band magnitudes of high redshift
 sources if
oscillations are present. We have shown that it is not
 possible to
use current data to obtain any further constraints on this
region. However, a future SNIa sample could yield better constraints.

\section*{Acknowledgments}
The authors would like to thank Ariel Goobar, Lars Bergstr\"om and
Eric Linder for useful discussions and Ching Wa-Yip for valuable help
with the SDSS Data Archive Server. Funding for the creation and
distribution of the SDSS Archive has been provided by the Alfred
P. Sloan Foundation, the Participating Institutions, the National
Aeronautics and Space Administration, the National Science
Foundation, the US Department of Energy, the Japanese Monbukagakusho,
and the Max Planck Society. The SDSS Web site is
http://www.sdss.org/. The Participating Institutions are the
University of Chicago, Fermilab, the Institute for Advanced Study,
the Japan Participation Group, the Johns Hopkins University, the Max
Planck Institute for Astronomy (MPIA), the Max Planck Institute for
Astrophysics (MPA), New Mexico State University, Princeton
University, the United States Naval Observatory, and the University
of Washington.

\end{document}